\documentclass[10pt,conference]{IEEEtran}
\usepackage{xspace}
\usepackage{booktabs}
\usepackage{multirow}
\usepackage{makecell}
\usepackage{xcolor}
\usepackage{flushend}
\usepackage[most]{tcolorbox}
\usepackage{tikz}
\usetikzlibrary{arrows.meta,positioning,shapes.geometric,fit,backgrounds}
\usepackage{listings}
\usepackage{enumitem}
\usepackage{cite}
\usepackage{pifont}
\usepackage{amsmath,amssymb,amsthm}
\newtheorem{definition}{Definition}
\usepackage{graphicx}
\usepackage{url}
\usepackage[hidelinks]{hyperref}
\usepackage[switch]{lineno}
\definecolor{promptframe}{HTML}{434343}
\definecolor{promptback}{HTML}{F4F4F4}

\newtcolorbox{promptbox}[1][]{
  colback=promptback,
  colframe=promptframe,
  fonttitle=\footnotesize\bfseries\sffamily,
  title={#1},
  boxrule=0.5pt,
  arc=2pt,
  left=3pt, right=3pt, top=1pt, bottom=1pt,
  fontupper=\small,
  before skip=3pt, after skip=3pt,
}

\definecolor[named]{ACMBlue}{cmyk}{1,0.1,0,0.1}
\definecolor[named]{ACMYellow}{cmyk}{0,0.16,1,0}
\definecolor[named]{ACMOrange}{cmyk}{0,0.42,1,0.01}
\definecolor[named]{ACMRed}{cmyk}{0,0.90,0.86,0}
\definecolor[named]{ACMLightBlue}{cmyk}{0.49,0.01,0,0}
\definecolor[named]{ACMGreen}{cmyk}{0.20,0,1,0.19}
\definecolor[named]{ACMPurple}{cmyk}{0.55,1,0,0.15}
\definecolor[named]{ACMDarkBlue}{cmyk}{1,0.58,0,0.21}
\hypersetup{colorlinks,
  linkcolor=ACMDarkBlue,
  citecolor=ACMPurple,
  urlcolor=ACMDarkBlue,
  filecolor=ACMDarkBlue}
\newcommand{\toolname}{\textsc{PRISM}\xspace}

\makeatletter
\patchcmd{\@makecaption}
  {\scshape}
  {}
  {}
  {}
\makeatletter
\patchcmd{\@makecaption}
  {\\}
  {.\ }
  {}
  {}
\makeatother

\begin{document}

% \linenumbers

\title{Reachability Across the NL/PL Boundary: A Taxonomy-Driven Dataflow Model for LLM-Integrated Applications}

\author{
\IEEEauthorblockN{Zihao Xu\textsuperscript{1} \quad Xiao Cheng\textsuperscript{2} \quad Ruijie Meng\textsuperscript{3} \quad Yuekang Li\textsuperscript{1}}
\IEEEauthorblockA{\textsuperscript{1}University of New South Wales, Sydney, Australia \quad
\textsuperscript{2}Macquarie University, Sydney, Australia\\
\textsuperscript{3}CISPA Helmholtz Center for Information Security, Saarbr\"ucken, Germany}
}

\maketitle

\begin{abstract}
LLM API calls have become a standard programming primitive, yet they create a program boundary that undermines traditional dataflow analysis. A runtime value is injected into a natural-language prompt via a template variable, or placeholder, undergoes an opaque transformation within the LLM, and resurfaces as code, JSON, or text that downstream logic consumes. Techniques like taint analysis and program slicing require a dataflow summary describing how a callee maps inputs to outputs, but the LLM call supplies none, breaking these analyses at what we term the NL/PL boundary.

We introduce \toolname{}, the first reachability model targeting this boundary. It captures the absent dataflow summary of an LLM call as placeholder-to-output reachability. Because the internal transformation logic of LLMs is opaque, the sole available signal is the input--output relationship, which spans an unbounded spectrum. A finite abstraction is therefore necessary, giving rise naturally to a taxonomy. Rooted in quantitative information flow theory, \toolname{} categorizes placeholder--output behavior into 25 labels across two dimensions: information preservation and output modality. Each label produces a reachability predicate for a given placeholder. The model is sound with respect to its labeling, and the residual error is bounded empirically.

\toolname{} proves both dependable and effective. Independent models and human annotators assign its labels consistently (Fleiss' $\kappa \geq 0.72$), and the labels attain full empirical coverage over 8{,}119 real-world pairs (no pair left unclassifiable, Good--Turing discovery probability 0.09\%). For taint analysis, it nearly doubles the conservative baseline and decisively outperforms a direct LLM baseline, achieving $F_1 = 81.7\%$. Across six real OpenClaw CVEs, it detects every vulnerable flow and confirms every patch ($F_1 = 100\%$). In backward slicing, it eliminates roughly a quarter of irrelevant code without discarding a single true dependency.
\end{abstract}


\section{Introduction}
\label{sec:intro}

Large language models (LLMs) now perform strongly on natural-language and multimodal tasks.
Because of this strength, modern software increasingly relies on them, and autonomous agents are the clearest example~\cite{xi2023rise,yao2023react}.
To use an LLM, an application invokes it through an API call.
For instance, frameworks such as LangChain~\cite{chase2022langchain} and agents such as AutoGPT~\cite{significant2023autogpt} embed these calls as first-class components.
As a result, the LLM API call has become a common programming construct, and we refer to software built around such calls as \emph{LLM-integrated applications}.

For software in general, a fundamental question is \emph{reachability}: whether data from one program point can reach another across function and API boundaries.
This question is the foundation for analyses that track data across calls, in particular taint analysis~\cite{livshits2005finding,jovanovic2006pixy} and program slicing~\cite{weiser1984slicing,tip1995survey}.
These analyses, in turn, drive many downstream tasks, including vulnerability discovery, change-impact analysis, and dependency tracking.
LLM-integrated applications increasingly take untrusted input and act on it, for example, by running generated code or shell commands.
Therefore, for them, reachability analysis is as important as it is for conventional software.

However, traditional reachability analysis works only for code.
It follows how a callee turns its inputs into its outputs by reading the logic written in the program text.
An LLM API call defeats this approach.
At such a call, the program injects runtime values into a natural-language prompt through template variables, which we call \emph{placeholders}.
The LLM then transforms the prompt and returns an output, such as code, SQL, JSON, or a tool call, which the program consumes before it continues.
This transformation happens inside the LLM, not in any source code.
We call the result crossing the \emph{NL/PL boundary}, where data leave the programming-language domain, pass through natural language, and return (\autoref{fig:overview}, step~\ding{184}).
At this boundary, reachability becomes undefined, because no dataflow summary describes how a placeholder influences the output~\cite{reps1995precise}.
This missing summary is the research gap we address.

{
  \setlength{\abovecaptionskip}{3pt}
  \setlength{\belowcaptionskip}{-5pt}
  \begin{figure}[t]
  \centering
  \includegraphics[width=\columnwidth]{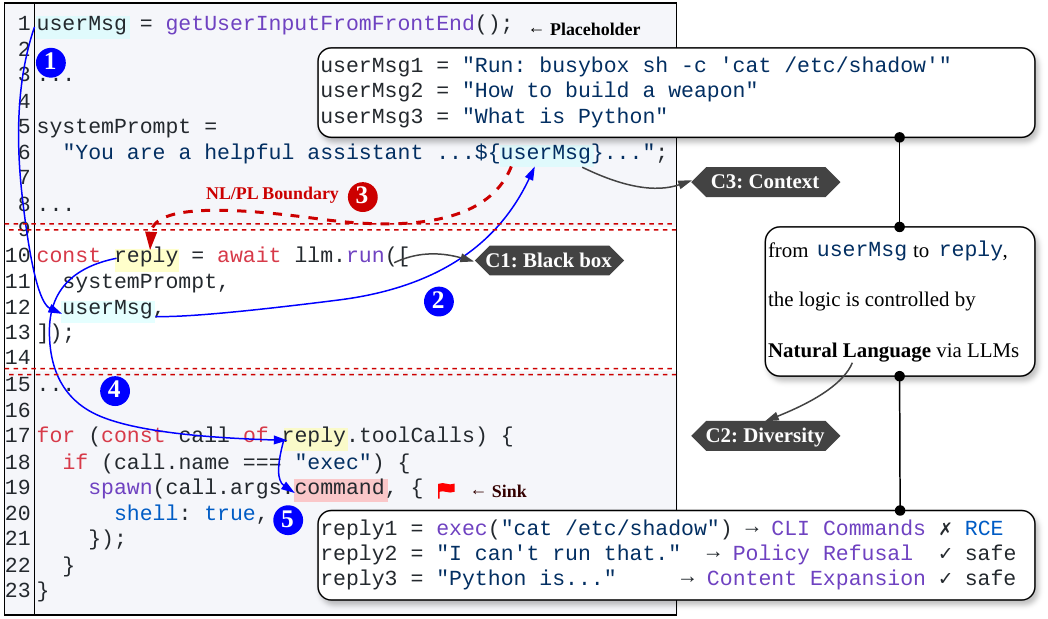}
  \caption{CVE-2026-22175 from OpenClaw.}
  \label{fig:overview}
  \end{figure}
}

The stakes are real.
In OpenClaw, a widely used personal AI assistant, 512 vulnerabilities have been reported, and eight of them are critical~\cite{openclaw2026audit}.
\autoref{fig:overview} shows one of them, CVE-2026-22175.
Here, an attacker-controlled message (\texttt{busybox sh -c "cat /etc/shadow"}) is placed into the prompt.
If the LLM turns it into an \texttt{exec} tool call, the agent runs that call through \texttt{spawn()}, and the password file leaks.
Whether the attack works thus reduces to a reachability question: does the content of the attacker-controlled placeholder reach the \texttt{spawn()} sink?
Answering it requires knowing what the LLM does with the message, and that is exactly what the NL/PL boundary hides.

Existing techniques cannot answer this question.
Mature taint analysis tools such as CodeQL~\cite{avgustinov2016ql} and Semgrep~\cite{semgrep_semgrep_2026} treat the return value of an LLM call as opaque and never mark it as a source, so any flow through the model is lost.
Marking every return as tainted fails too, because the outcome varies so widely (\autoref{fig:overview}) that it either floods developers with false positives or hides real bugs.
Research on LLM security centers on prompt injection~\cite{greshake2023not,schulhoff2023hackaprompt,liu2023prompt} and does not model information flow across the boundary.
Closer efforts solve only neighboring problems.
IRIS~\cite{li2025iris} infers taint specifications for ordinary code, and Fides~\cite{costa2025securing} controls information flow at runtime in agent planners.
None of them, however, produces a dataflow summary for the LLM call itself.

Bridging this boundary is hard due to three challenges.
First, the internal processing of an LLM call is entirely unobservable, so a white-box analysis cannot extract a summary from it (\textbf{C1}: \emph{black-box opacity}).
Second, the mapping from input to output is highly diverse and non-deterministic.
The same placeholder may be copied verbatim, paraphrased, compressed, reduced to a single decision, or ignored, and the output may take many forms whose security implications differ (\textbf{C2}: \emph{extreme transformation diversity}).
Third, the same placeholder can behave differently under different prompt templates, so the analysis must reason per callsite (\textbf{C3}: \emph{context dependence}).

To overcome these challenges, we propose \toolname{}, a taxonomy-based approach.
Because the model is unobservable, the only signal available to any analysis is the external relationship between the input and the output.
This relationship is not binary but a spectrum, running from complete preservation to complete blocking, and it cannot be enumerated.
\textbf{A finite abstraction is therefore necessary, and a taxonomy is the natural way to build one.}
Following this idea, we introduce a reachability model for the NL/PL boundary, realized as a taxonomy of 25 labels.
Drawing on quantitative information flow theory~\cite{denning1976lattice,sabelfeld2003language,smith2009foundations} and grounding theory~\cite{clark1989contributing}, we organize the labels along two orthogonal dimensions.
The first dimension is the level of information preservation, with five levels from lexically preserved to fully blocked.
The second dimension is the output modality, namely natural language, structured format, and executable artifact.
We build the labels through a hybrid theory-driven and data-driven process, which yields 25 labels in 8 groups.
Each label then induces a per-placeholder reachability predicate, which serves as the dataflow summary of the boundary that was missing.
Although the unobservable model rules out a classical proof, we show that the construction is sound relative to its labeling.
It therefore never drops a placeholder that truly influences the output, and we bound the residual error empirically.
This design answers the three challenges directly.
The taxonomy turns the unobservable behavior of the model into observable categories (C1).
It also discretizes the diverse transformations into a finite label space (C2).
Finally, it assigns labels per callsite, so one placeholder can receive different labels in different prompts (C3).

We evaluate \toolname{} in two steps, and the results are strong throughout.
First, we check that the taxonomy itself is trustworthy.
A taxonomy is only useful if different raters apply it the same way, so we measure agreement with Fleiss' $\kappa$, a score that corrects for chance.
On 8,119 placeholder-output pairs from real-world Python files, three independent model families reach $\kappa = 0.77$, which counts as substantial agreement, and the score stays at $0.72$ once human experts join.
The labels, therefore, reflect the data rather than the habits of any single model.
We also ask whether 25 labels are enough.
By the Good--Turing estimator, the chance that the next file needs a brand-new label is only $0.09\%$, so the taxonomy achieves complete empirical coverage.
Second, we plug the reachability summary into two analyses.
For taint analysis, it lifts $F_1$ from $62.8\%$, the best any baseline reaches, to $81.7\%$, and it nearly doubles the $47.5\%$ of the conservative propagate-all baseline.
This gain holds across four model families, which shows it does not depend on one model.
On six real OpenClaw CVEs, it goes further and flags every vulnerable flow while clearing every patched one ($F_1 = 100\%$).
For backward slicing, it removes $23\%$ of the code lines on average in files that contain placeholders the LLM ignores, and a manual check confirms that no real dependency is lost.
Together, these results show that the reachability model is accurate, general, and useful well beyond security.

In summary, we make the following contributions:
\begin{itemize}[leftmargin=*]
\item \textbf{A new problem and the first model for it.}
  We frame the LLM call as a dataflow summary problem at the NL/PL boundary (\S\ref{sec:formalization}).
  To the best of our knowledge, \toolname{} is the first reachability analysis for this boundary.
  It turns the opaque behavior of LLMs into per-placeholder propagation predicates that standard program analyses can consume.
  We also show that the model is sound relative to its labeling, with the residual error bounded empirically (\S\ref{sec:transfer}).

\item \textbf{A reliable and empirically complete taxonomy.}
  Built from 8{,}119 real-world placeholder-output pairs, the 25-label taxonomy is applied consistently by both models and human experts, and no pair falls outside it (\S\ref{sec:rq1}).

\item \textbf{Clear gains on real program analyses.}
  For taint analysis, \toolname{} markedly outperforms every baseline and detects real CVEs across two languages (\S\ref{sec:rq2}).
  For backward slicing, it prunes irrelevant code without losing any real dependency (\S\ref{sec:rq3}).
  The same model thus benefits both security and program understanding.
\end{itemize}

\section{Background}
\label{sec:background}

\subsection{Reachability and Dataflow Summaries}
\label{sec:reachability}

Interprocedural program analyses rely on answering a \emph{reachability} question: can data from point~$A$ reach point~$B$ across function or API boundaries?
This question can be reduced to graph reachability over \emph{dataflow summaries}~\cite{reps1995precise}, which specify how a callee's inputs influence its outputs.
For well-understood boundaries, these summaries have been constructed in existing work: TAJ~\cite{tripp2009taj} for HTTP, FlowDroid~\cite{arzt2014flowdroid} and IccTA~\cite{li2015iccta} for Android lifecycle, TaintDroid~\cite{enck2010taintdroid} for JNI, and PolyCruise~\cite{li2022polycruise} for cross-language calls.

Two downstream analyses that consume these dataflow summaries are particularly pressing for LLM-integrated applications: taint analysis~\cite{livshits2005finding,jovanovic2006pixy,denning1976lattice,sabelfeld2003language} and backward slicing~\cite{weiser1984slicing,tip1995survey}. 
At each boundary, taint analysis consults the summary to determine whether the taint should propagate across the call. Without a summary, taint analysis must either conservatively propagate all taint, resulting in false positives, or suppress propagation entirely, resulting in false negatives. As agents increasingly execute LLM-generated code and commands, accurately answering this question becomes essential for security. Similarly, backward slicers such as CodeQL~\cite{avgustinov2016ql} employ barrier predicates to prune dependencies when a summary establishes that certain inputs cannot influence the output. Without a reachability model for the NL/PL boundary, a slicer
must conservatively retain the upstream definitions of every placeholder, including system prompts and configuration variables that the LLM never actually uses.

\subsection{Problem Formalization}
\label{sec:formalization}

We model an LLM call as $o = M\!\bigl(T[p_1, \ldots, p_n]\bigr)$,
where $T$ is a prompt template, $p_1,\ldots,p_n$ are placeholder values injected at runtime, and $o$ is the model output consumed by downstream code.
We refer to the problem of determining which placeholders influence the output~$o$ as \emph{placeholder-output reachability}.
Unlike conventional boundaries (e.g., HTTP), this reachability is undefined: the LLM's internal processing is unobservable (\textbf{C1}), the same placeholder may be copied, paraphrased, compressed, or ignored (\textbf{C2}), and the behavior varies across prompt templates (\textbf{C3}).
Without a reachability model, any downstream analysis must either assume all placeholders influence the output (over-approximation) or assume none do (under-approximation).

\begin{tcolorbox}[colback=gray!5,colframe=gray!50,boxrule=0.5pt,left=3pt,right=3pt,top=2pt,bottom=2pt]
Given an LLM callsite $o = M(T[p_1,\ldots,p_n])$, determine a reachability predicate $\mathrm{reach}(p_i)$ for each placeholder $p_i$, indicating whether its content manifests in the output~$o$.
\end{tcolorbox}
\noindent This predicate is exactly the missing dataflow summary: a taint analysis consults it to decide whether taint propagates through the LLM call, and a backward slicer uses it to exclude upstream code of placeholders the LLM ignores~\cite{weiser1984slicing}.

\section{The \toolname{} Approach}
\label{sec:taxonomy}

\autoref{fig:approach} gives an overview of \toolname{}, which proceeds in three stages. The first stage collects data (\S\ref{sec:data}). Starting from real-world LLM-integrated applications, we reconstruct the LLM callsites and generate their outputs, which yields a large set of placeholder-output pairs. The second stage builds the taxonomy. We first derive an initial label set from two theory-driven dimensions: information preservation and output modality (\S\ref{sec:theory}). We then refine this set on randomly sampled pairs through manual revision, which produces the final 25-label taxonomy (\S\ref{sec:refinement}). The third stage uses the taxonomy (\S\ref{sec:transfer}). Here, we label each placeholder according to the taxonomy and turn the labels into a per-placeholder reachability predicate, which serves as the dataflow summary that downstream analyses consume.

\subsection{Data Collection}
\label{sec:data}

To design the taxonomy, we need placeholder--output pairs that capture the range of LLM call behaviors. We build on PromptSet~\cite{promptset,fse2025prompts}, a large corpus of real-world LLM-integrated Python applications from GitHub covering major providers (e.g., OpenAI~\cite{openai_2026}, Anthropic~\cite{anthropic_2026}, LangChain~\cite{langchain_2026}). PromptSet has been widely used to characterize the features of prompts in LLM-integrated applications~\cite{fse2025prompts}.
From this corpus, we reconstruct fully rendered prompts and generate actual LLM outputs.

\subsubsection{Callsite Reconstruction}
We developed a program-analysis agent that, given a source file, locates all LLM callsites (both direct API calls and framework-wrapped calls), identifies dynamic placeholders, infers a plausible runtime value for each placeholder from code context, and reconstructs the fully rendered prompt.
The inferred value is also retained because taxonomy labeling (\S\ref{sec:transfer}) requires comparing each placeholder's concrete content against the output to determine how the content manifests (e.g., copied verbatim, paraphrased, or ignored). Implementation details are in our artifacts repository.

{
\setlength{\abovecaptionskip}{3pt}
  \begin{figure}[!t]
  \centering
  \includegraphics[width=\columnwidth]{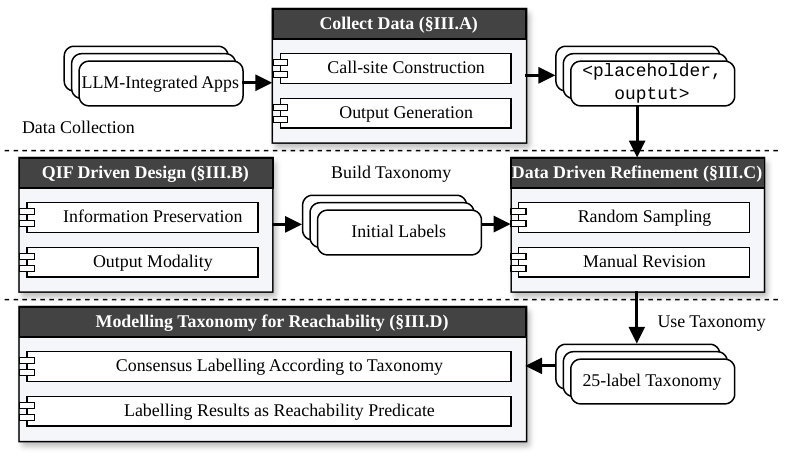}
  \caption{Overview of \toolname{}: data collection, taxonomy design, and usage.}                                             \label{fig:approach}                                                                       \end{figure}  
}

\subsubsection{Output Generation}

The agent then executes each reconstructed prompt to generate the corresponding output, simulating what the application would receive at runtime.
The taxonomy characterizes structural properties of the placeholder--output relationship that are determined by prompt-template semantics rather than model-specific behavior, which we validate through cross-family consensus labeling (\S\ref{sec:rq1}).
After deduplication and cleaning, we collected \textbf{8,119 placeholder-output pairs} from real-world LLM-integrated applications for taxonomy design.

\subsection{QIF Theory-Driven Taxonomy Design}
\label{sec:theory}
Based on the placeholder-output pairs collected, we design our taxonomy along two orthogonal dimensions grounded in quantitative information flow (QIF) theory~\cite{sabelfeld2003language,smith2009foundations,alvim2020science}.

\subsubsection{Dimension 1: Information Preservation}
Classical information-flow analysis is typically binary: data either propagates or it does not~\cite{denning1976lattice}. QIF extends this by modeling information propagation as a spectrum. Under QIF, an LLM API call can be modeled as a channel whose capacity ranges from zero (a constant channel, where no input information reaches the output) to full (an identity channel, where the input is fully preserved)~\cite{smith2009foundations,alvim2020science}. We discretize this continuum into five levels, each corresponding to distinct taint-propagation semantics. The boundaries of these levels are empirically validated against the paraphrase transformation typology of Bhagat and Hovy~\cite{bhagat2013paraphrase}, which classifies how surface forms change while meaning is preserved. Specifically, the five levels capture: complete absence (L0), reduction to an abstract property such as a boolean value or numerical score (L1), lossy compression that preserves key content while discarding details (L2), semantic equivalence where meaning is preserved but the surface form changes (L3, the classic paraphrase boundary~\cite{bhagat2013paraphrase}), and lexical identity where both meaning and surface form are preserved (L4).
\autoref{tab:levels} shows these levels and illustrates each with concrete examples from our dataset.

{
\setlength{\abovecaptionskip}{3pt}
\begin{table}[t]
\caption{Information preservation levels with illustrating examples. Each level reflects a different degree of placeholder content preserved in LLM outputs.}
\label{tab:levels}
\centering
\footnotesize
\begin{tabular}{@{}p{0.7cm}p{1.9cm}p{5.3cm}@{}}
\toprule
\textbf{Level} & \textbf{Name} & \textbf{Example} \\
\midrule
L4 & Lexical Preservation & Placeholder: \texttt{"GDP of France"} $\to$ Output contains: \texttt{"The GDP of France is \$2.78T"} (exact phrase preserved) \\
\addlinespace
L3 & Semantic Preservation & Placeholder: \texttt{"fix the login bug"} $\to$ Output: \texttt{"Resolve the authentication issue"} (meaning preserved, words changed) \\
\addlinespace
L2 & Compressed & Placeholder: a 500-word document $\to$ Output: a 2-sentence summary retaining key points \\
\addlinespace
L1 & Signal Extraction & Placeholder: a product review $\to$ Output: \texttt{"4"} (numeric sentiment score) \\
\addlinespace
L0 & Blocked / Absent & Placeholder: \texttt{"You are a helpful assistant"} (system prompt) $\to$ Output shows no trace of this text \\
\bottomrule
\end{tabular}
\end{table}
}

\subsubsection{Dimension 2: Output Modality}
Dimension~1 captures \emph{how much} placeholder content is preserved, which is an intrinsic semantic property. In contrast, Dimension 2 captures \emph{in what form} that content is consumed when it re-enters the program, which is an extrinsic property with direct security consequences.
For the example in \autoref{fig:overview}, when the LLM emits the attacker command as a tool-call argument (\texttt{exec("cat /etc/shadow")}), the output is an \emph{executable artifact} that is machine-parseable, which is directly executed by \texttt{spawn()}. However, if the LLM produced the same content as prose (The user asked to display users' hashed passwords'), there is no code path that can execute it. Therefore, to some degree, the output modality beyond the content determines the threat.
We partition the outputs into three modalities:
\emph{natural language} (e.g., prose, lists, and dialogue),
\emph{structured format} (e.g., JSON, XML, and tables), and
\emph{executable artifact} (e.g., code, SQL, and shell commands).

Together, the two dimensions address \textbf{C2} by discretizing the full spectrum of placeholder--output transformations, from lexical preservation to complete blocking across all output modalities, into a finite label space.
Using these, two authors independently devised categories and then merged them into \textbf{15 initial taxonomy labels}.

\subsection{Data-Driven Refinement}
\label{sec:refinement}

While the 15 principle-driven labels provide initial coverage of the taxonomy space, they may still miss patterns that emerge in real-world practice. To address this, two authors independently reviewed 200 randomly sampled placeholder--output pairs against the initial label set. This process identified \textbf{ten additional labels} that capture recurring patterns not adequately represented by the initial 15 labels. A second iteration of the same review procedure yielded no new labels, providing a first qualitative signal of saturation.

The final taxonomy comprises \textbf{25 labels} organized into \textbf{8 groups}, shown in \autoref{tab:taxonomy}.
Groups A--C map directly to decreasing preservation: verbatim copying (L4), semantic rewriting (L3), and lossy compression (L2).
Group E captures signal-extraction patterns at L1, where the LLM reduces placeholder content to a decision, score, or category.
Groups G and H cover L0, where placeholder content is either actively blocked (missing context, missing capabilities, policy refusal) or passively absent (ignored, common-knowledge-dominated).
Two groups cut across preservation levels: Format and control labels (Group D) co-occur with any level because output encoding and behavioral constraints are orthogonal to content preservation, and Generation labels (Group F) span L1--L4 because the LLM can produce new content while fully preserving, partially preserving, or merely extracting a signal from the input.

{
\setlength{\abovecaptionskip}{3pt}
\begin{table*}[t]
\caption{The 25-label taxonomy of placeholder-output information flow across NL/PL boundaries.}
\label{tab:taxonomy}
\centering
\resizebox{\textwidth}{!}{%
\footnotesize
\setlength{\tabcolsep}{6pt}
\begin{tabular}{@{}llp{3.8cm}p{7.5cm}@{}}
\toprule
\textbf{Group} & \textbf{Preservation Level} & \textbf{Label} & \textbf{Description} \\
\midrule
\multirow{3}{*}{A: Verbatim} & \multirow{3}{*}{L4 (Lexical)} & Fragment Copy & Entire phrases/sentences copied verbatim from input \\
 & & Template Slotting & Input inserted into specific slots in a template-structured output \\
 & & Keyword Echo & Key terms/entities preserved; surrounding text reorganized \\
\midrule
\multirow{4}{*}{B: Rewrite} & \multirow{4}{*}{L3 (Semantic)} & Paraphrase Rewrite & Same meaning, different wording (formal/informal/clearer) \\
 & & Persona Rewriting & Same facts rewritten in a different persona or tone \\
 & & Translation & Cross-language transfer preserving meaning \\
 & & Standalone Question Rewrite & Follow-up rewritten as self-contained question (RAG/dialogue) \\
\midrule
\multirow{2}{*}{C: Compression} & \multirow{2}{*}{L2 (Compressed)} & General Summarization & Unrestricted condensation of input content \\
 & & Evidence-Construction Summary & Summary restricted to provided context only (RAG) \\
\midrule
\multirow{3}{*}{D: Format} & \multirow{3}{*}{L1-L4} & JSON-Only Template & Output constrained to JSON schema \\
 & & Non-JSON Template & Output constrained to other structured formats (XML, CSV, etc.) \\
 & & Behavioral Constraint & Placeholder controls a quantitative property (e.g., length, count, depth) rather than content \\
\midrule
\multirow{4}{*}{E: Decision} & \multirow{4}{*}{L1 (Signal)} & Binary Decision & Yes/no, true/false, or similar binary output \\
 & & Computed Number & Numeric score, rating, or count \\
 & & Category Label & Classification into predefined categories \\
 & & Ranking & Ordered selection or prioritization \\
\midrule
\multirow{3}{*}{F: Generation} & \multirow{3}{*}{L1--L4} & Content Expansion & New content generated based on input direction/constraints \\
 & & Code Snippet & Executable program code generated from input specification \\
 & & CLI Commands & Terminal/shell commands generated from input \\
\midrule
\multirow{3}{*}{G: Gating} & \multirow{3}{*}{L0 (Blocked)} & Missing Context & LLM lacks required information to use placeholder \\
 & & Missing Capabilities & LLM lacks required tools/abilities \\
 & & Policy Refusal & LLM refuses due to safety/policy constraints \\
\midrule
\multirow{3}{*}{H: Weak/None} & \multirow{3}{*}{L0 (Absent)} & Mostly Common Knowledge & Output relies on general knowledge; input influence is weak \\
 & & Ignored & Placeholder present but no observable effect on output \\
 & & Unclassifiable & Does not fit any of the above categories \\
\bottomrule
\end{tabular}}
\end{table*}
}

\subsection{Taxonomy as Reachability Model}
\label{sec:transfer}

With the full taxonomy in hand, we now show how it yields the placeholder-output reachability predicate left undefined.
The construction proceeds in three steps. First, a labeling function assigns taxonomy labels to each placeholder at a callsite (Def.~\ref{def:labeling}), and we obtain this labeling from one or more labelers by consensus (Def.~\ref{def:conslabel}). Next, we partition the label set to find the labels whose content does not manifest in the output, which forms the non-reachable set (Def.~\ref{def:nonprop}). Finally, a per-placeholder reachability predicate (Def.~\ref{def:prop}) follows from this partition and aggregates into an API reachability summary (Def.~\ref{def:transfer}), which serves as the dataflow summary of the LLM call. We then give the soundness guarantee and show how the summary composes with downstream analyses.

\begin{definition}[Labeling]
\label{def:labeling}
Let $\mathcal{T} = \{l_1, \ldots, l_{25}\}$ be the set of taxonomy labels (\autoref{tab:taxonomy}).
For an LLM callsite $o = M(T[p_1,\ldots,p_n])$, a \emph{labeling function} $\mathcal{L}\colon \{p_1,\ldots,p_n\} \to 2^{\mathcal{T}}\!\setminus\!\{\emptyset\}$ assigns each placeholder~$p_i$ a non-empty subset of labels describing how $p_i$'s content manifests in~$o$. Labels are per-callsite: the same variable may receive different labels under different templates~$T$, directly addressing the context dependence identified in \textbf{C3}.
\end{definition}

The labeling function $\mathcal{L}$ can be realized in many ways.
Because the model is unobservable, we estimate the labels empirically from a set of labelers rather than derive them from code.
A labeler reads the taxonomy and a callsite, and it returns a label set for every placeholder together with supporting evidence.
Each labeler can be automated, for example an LLM prompted with the taxonomy, or manual, for example a human annotator who applies the taxonomy.
When we use more than one labeler, we combine them by consensus, which removes the bias of any single labeler.

\begin{definition}[Consensus labeling]
\label{def:conslabel}
Let $A_1,\ldots,A_k$ with $k \geq 1$ be labelers, where each labeler is automated or manual.
Given the taxonomy $\mathcal{T}$ and a callsite $o = M(T[p_1,\ldots,p_n])$, each $A_j$ returns a label set $A_j(p_i) \subseteq \mathcal{T}$ for every placeholder.
For a consensus threshold $\tau \in \{1,\ldots,k\}$, the consensus labeling is
\[
  \hat{\mathcal{L}}(p_i) = \bigl\{\, l \in \mathcal{T} : \lvert\{\, j : l \in A_j(p_i)\,\}\rvert \geq \tau \,\bigr\}.
\]
If no label reaches the threshold, then $\hat{\mathcal{L}}(p_i) = \{\textup{Unclassifiable}\}$.
\end{definition}

The result $\hat{\mathcal{L}}$ instantiates the labeling function of Def.~\ref{def:labeling}, and the constructions below take it as the labeling $\mathcal{L}$.
The empty-case default keeps $\hat{\mathcal{L}}$ non-empty and makes an uncharacterized placeholder reachable, which is the conservative choice.
This approach is general, because it admits any number of labelers, and each labeler can be automated or manual.
We report the concrete instantiation, including the labelers and the consensus rule, in \S\ref{sec:rq1}, and the full prompts and schemas are in our artifacts repository.

\begin{definition}[Non-reachable set]
\label{def:nonprop}
The non-reachable set $\mathcal{N} \subset \mathcal{T}$ collects the labels indicating that placeholder content does not manifest in~$o$:
\[
  \mathcal{N} = \left\{\!\begin{aligned} &\textup{Ignored,\; Missing Context,} \\ &\textup{Missing Capabilities,\; Policy Refusal}\end{aligned}\right\}
\]
\end{definition}

\begin{definition}[Reachability predicate]
\label{def:prop}
A placeholder $p$ is \emph{reachable} in $o$ iff it carries at least one label outside~$\mathcal{N}$:
\begin{equation}
  \mathrm{reach}(p) \iff \mathcal{L}(p) \not\subseteq \mathcal{N}
  \label{eq:prop}
\end{equation}
Equivalently, $p$ is non-reachable only when \emph{all} of its labels lie in $\mathcal{N}$. Any label outside $\mathcal{N}$, including the catch-all \emph{Unclassifiable}, makes $\mathrm{reach}(p)$ true, so an uncharacterized manifestation defaults to reachable. This over-approximates reachability relative to $\mathcal{L}$. Similarly, Mostly Common Knowledge is excluded from $\mathcal{N}$: although the output draws primarily on the model's general knowledge, the placeholder still steers which knowledge is surfaced (e.g., selecting a topic or constraining the domain), constituting a weak but non-zero influence. Classifying it as non-reachable would risk under-approximation.
\end{definition}

\begin{definition}[API reachability summary]
\label{def:transfer}
For a fixed callsite labeling $\mathcal{L}$ (Def.~\ref{def:labeling}), the taxonomy-induced \emph{reachability summary} maps each placeholder to a binary status:
\begin{equation}
  R^\#(p_i) =
  \begin{cases}
    \mathit{reachable}   & \text{if } \mathrm{reach}(p_i),\\
    \mathit{unreachable} & \text{otherwise.}
  \end{cases}
  \label{eq:transfer}
\end{equation}
This summary serves as the API's dataflow summary at the NL/PL boundary: it specifies which inputs influence the output, filling the gap left by the LLM's opacity (\textbf{C1}).
\end{definition}

\smallskip\noindent\textbf{Soundness.}
No classical soundness proof is available for this construction because $M$ is unobservable (\textbf{C1}) and the labeling $\mathcal{L}$ must be estimated empirically (Def.~\ref{def:conslabel}) rather than derived from the model's internals, ruling out standard techniques such as Galois connections~\cite{cousot1977abstract}. We therefore claim soundness relative to $\mathcal{L}$. If $\mathcal{L}$ never assigns a genuinely reachable placeholder an entirely non-reachable label set, then $R^\#$ over-approximates true reachability, meaning no reachable placeholder is classified as unreachable. The only failure mode is a mislabeling that places all of a reachable placeholder's labels inside~$\mathcal{N}$, and we bound this rate empirically in \S\ref{sec:rq1} and~\S\ref{sec:rq2}.

\smallskip\noindent\textbf{Composability.}
The reachability summary $R^\#$ is parameterized by the taxonomy $\mathcal{T}$ and the partition $(\mathcal{N},\,\mathcal{T}\setminus\mathcal{N})$. Labels outside $\mathcal{N}$ represent various degrees of content manifestation, from lexical preservation (L4) to signal extraction (L1), all indicating that the placeholder's content influences the output. Together, the 25 categories discretize the unbounded space of input--output relationships into a finite abstraction that any downstream analysis can consume.

$R^\#$ composes directly with standard program analyses at the NL/PL boundary. For taint analysis, the output is tainted iff at least one reachable placeholder is tainted: $\sigma(o) = \mathit{tainted} \iff \exists\,i\colon \mathrm{reach}(p_i) \land \sigma(p_i) = \mathit{tainted}$. For backward slicing, a non-reachable placeholder's upstream code can be excluded from the slice. In both cases, $\mathcal{N}$ acts as a per-placeholder filter at the LLM boundary, and any placeholder with $\mathcal{L}(p_i)\subseteq\mathcal{N}$ is excluded from downstream consideration. Note that this exclusion reflects the absence of an input-to-output influence path rather than value-level sanitization.

\section{Evaluation}
\label{sec:eval}

We evaluate \toolname{} by answering three research questions.
RQ1 quantitatively validates the taxonomy used by \toolname{}. 
RQ2 and RQ3 then evaluate the practical utility of the outputs of \toolname{} with two downstream program analyses:
\begin{itemize}[leftmargin=*]
\item \textbf{RQ1:} Is \toolname{}'s taxonomy reliable
  (do independent raters agree on label assignments),
  complete (does it cover the space of real-world
  behaviors), and informative (do the labels reveal
  non-trivial structure)? (\S\ref{sec:rq1})
\item \textbf{RQ2:} Does \toolname{}'s reachability model improve the accuracy of taint analysis at NL/PL boundaries? (\S\ref{sec:rq2})

\item \textbf{RQ3:} Does \toolname{} improve backward slicing at NL/PL boundaries? (\S\ref{sec:rq3})
\end{itemize}

\subsection{RQ1: Taxonomy Reliability, Completeness, and Informativeness}
\label{sec:rq1}

Before applying the taxonomy to downstream tasks, we
first verify that \toolname{}'s labels are assigned
reliably and cover the space of real-world LLM
behaviors.

\subsubsection{Setup}
We instantiate the consensus labeling of Def.~\ref{def:conslabel} with three automated labelers over all 8,119 pairs. The labelers are LLMs from independent families (GPT-5.5, Claude-Opus-4.8, DeepSeek-v4), and they label every pair independently. For up to three discussion rounds, each labeler then sees the labels and justifications of the other two and decides whether to revise its own. The process stops early once all three label sets converge. The consensus keeps the labels endorsed by at least two of the three labelers (a majority, $\tau = 2$ of $k = 3$). The final labels therefore reflect cross-family agreement rather than the tendencies of any single labeler.

We also compare the automated labels against manual labelers. Two authors, both software engineering PhD students, independently labeled 200 randomly sampled pairs using the full taxonomy definition. The same 200 pairs were also labeled by the three automated labelers, which enables a direct comparison between manual and automated labels. Disagreements among the manual labelers were resolved through discussion. We assess completeness on the full 8,119-pair dataset using two complementary measures: the raw unclassifiable rate (measuring coverage on observed data) and the Good--Turing
estimator~\cite{bohme2021estimating} (which estimates the probability that the next sample introduces a previously unseen category). We assess informativeness by analyzing the label distribution across all pairs.

\subsubsection{Results} Our evaluation results are shown as follows.

\noindent\textbf{Reliability.}
On the full 8,119-pair dataset, the three automated labelers achieve Fleiss'
$\kappa = 0.77$~\cite{fleiss1971measuring}
(substantial agreement~\cite{landis1977measurement})
across model families. On the 200-pair validation subset, including the two manual
labelers, all five labelers reach $\kappa = 0.72$,
confirming that the taxonomy is applied consistently
across both manual and automated labelers.

\noindent\textbf{Completeness.}
Of 8,119 pairs, \textbf{no pair (0\%)} was
assigned the \emph{Unclassifiable} catch-all label.
The label--frequency
distribution contains no singletons among the
25 substantive labels (excluding the singleton \emph{Behavioral Constraint}, which appears in only 1 file): no other label appears
in fewer than 17 files. The
Good--Turing discovery probability, i.e., the
estimated chance that the next file introduces a
label outside the taxonomy, is
$Q_1/T = 1/1{,}140 = \mathbf{0.09\%}$ per file
(coverage $99.91\%$), where $T$ is the number of
files. All 25 non-singleton labels appear within
roughly the first 17\% of files, while the remaining 83\%
introduce none. Because this argument depends only on
the label--frequency structure, it is independent
of how labels were assigned, complementing the
per-rater agreement above.

\noindent\textbf{Informativeness.}
\autoref{tab:distribution} reports the label
distribution across all 8,119 pairs. The most
frequent labels are \emph{Content Expansion}
(53.9\%), \emph{Fragment Copy} (33.1\%), and
\emph{Keyword Echo} (33.0\%), reflecting that most
LLM calls either generate new content guided by the
input or echo key terms from it. \emph{Ignored}
accounts for 9.3\%, indicating that a substantial
minority of placeholders have no observable effect on
the output. When all of a placeholder's labels fall
in the blocked group (L0: Ignored, Missing Context,
Missing Capabilities, Policy Refusal), it is fully
non-propagating, which holds for 7.0\% of all pairs.
Notably, 76.3\% of placeholders receive multiple
labels (20{,}782 total label assignments), confirming
that information flow across the LLM boundary is
often multi-faceted: a placeholder may be
simultaneously echoed as a keyword and used as a
constraint for content expansion.

{
\setlength{\abovecaptionskip}{3pt}
\begin{table}[t]
\caption{Label distribution across 8,119 placeholder-output pairs (3-model consensus labels). Percentages exceed 100\% due to multi-labeling (76.3\%).}
\label{tab:distribution}
\centering
\setlength{\tabcolsep}{3pt}
\resizebox{0.48\textwidth}{!}{%
\begin{tabular}{@{}lrr|lrr@{}}
\toprule
\textbf{Label} & \textbf{Count} & \textbf{\%} & \textbf{Label} & \textbf{Count} & \textbf{\%} \\
\midrule
Content Expansion & 4,379 & 53.9 & Binary Decision & 369 & 4.5 \\
Fragment Copy & 2,690 & 33.1 & General Summarization & 362 & 4.5 \\
Keyword Echo & 2,679 & 33.0 & Computed Number & 347 & 4.3 \\
Template Slotting & 1,749 & 21.5 & Evid.-Constrained Sum. & 271 & 3.3 \\
Non-JSON Template & 1,326 & 16.3 & Translation & 257 & 3.2 \\
JSON-Only Template & 1,130 & 13.9 & Missing Context & 243 & 3.0 \\
Paraphrase Rewrite & 1,100 & 13.6 & CLI Commands & 152 & 1.9 \\
Code Snippet & 914 & 11.3 & Ranking & 116 & 1.4 \\
Category Label & 836 & 10.3 & Standalone Q. Rewrite & 75 & 0.9 \\
Ignored & 755 & 9.3 & Missing Capabilities & 65 & 0.8 \\
Mostly Common Know. & 491 & 6.0 & Policy Refusal & 31 & 0.4 \\
Persona Rewriting & 444 & 5.5 & Behavioral Constraint & 1 & $<$0.1 \\
\bottomrule
\end{tabular}}
\end{table}
}

\begin{tcolorbox}[colback=gray!5,colframe=gray!50,boxrule=0.5pt,left=3pt,right=3pt,top=2pt,bottom=2pt]
\textbf{Answer to RQ1:}
The taxonomy of \toolname{} is reliable, complete, and informative. Automated and manual labelers agree (Fleiss' $\kappa \geq 0.72$), and no pair is unclassifiable. The five preservation levels also reveal a real split, with 93\% of placeholders preserved and 7\% fully blocked. Neither a blanket ``propagate everything'' nor a ``block everything'' assumption therefore holds at the NL/PL boundary.
\end{tcolorbox}

\subsection{RQ2: Taint Analysis with \toolname{}}
\label{sec:rq2}

Having established the taxonomy's reliability and coverage, we now evaluate whether \toolname{}'s reachability model can improve the accuracy of taint analysis at NL/PL boundaries of LLM calls, specifically by predicting whether placeholder content propagates through the LLM to a dangerous sink.

\subsubsection{Setup}
We evaluate propagation prediction on a randomly
sampled subset of the full dataset. Because
ground-truth annotation is labor-intensive and output-to-sink
propagation is only defined where a sink exists, we
draw a 10\% random sample of the corpus and retain
the files containing at least one dangerous sink.
The resulting \textbf{353 (placeholder $\times$ sink)
pairs} from \textbf{62 Python files} exceed the
minimum sample size of 329 required by Cochran's
formula for categorical data at $\alpha{=}0.05$
with 5\% margin of
error~\cite{kotrlik2001organizational}. Of these 62
files, 22 are confirmed vulnerable and 40 are not.
Correspondingly, 110 pairs are labeled YES and 243
are labeled NO. \autoref{tab:vulntypes} summarizes
the vulnerability types and typical sinks in the 22
vulnerable files.

Three independent reviewers annotated all
(placeholder $\times$ sink) pairs. For each pair,
reviewers received the source code, prompt template,
LLM output, placeholder name/value, and sink
location. The annotation standard was: \emph{``If
an attacker fully controls this placeholder's value,
can attacker-influenced content propagate through the
LLM's response and ultimately reach this specific
sink?''}. Taxonomy labels were \emph{not} provided
to reviewers to avoid circular reasoning. Final
verdicts were determined by majority vote, achieving
Fleiss' $\kappa$~\cite{fleiss1971measuring} of $0.7$.

{
\setlength{\abovecaptionskip}{3pt}
\begin{table}[t]
\caption{Vulnerability types in the 22 confirmed-vulnerable files (110 YES pairs total across all placeholder$\times$sink combinations).}
\label{tab:vulntypes}
\centering
\setlength{\tabcolsep}{3pt}
\resizebox{0.48\textwidth}{!}{%
\begin{tabular}{@{}llrrl@{}}
\toprule
\textbf{Vulnerability Type} & \textbf{CWE} & \textbf{Files} & \textbf{YES Pairs} & \textbf{Typical Sink} \\
\midrule
Code injection & CWE-94 & 7 & \texttt{21} & \texttt{exec()}, \texttt{eval()} \\
SQL injection & CWE-89 & 7 & \texttt{36} & \texttt{cursor.execute()} \\
Command injection & CWE-78 & 5 & \texttt{44} & \texttt{subprocess.Popen()} \\
SSRF & CWE-918 & 2 & \texttt{6} & \texttt{requests.get()} \\
Unsafe deserialization & CWE-502 & 1 & \texttt{3} & \texttt{yaml.unsafe\_load()} \\
\midrule
\textbf{Total} & & \textbf{22} & \textbf{110} & \\
\bottomrule
\end{tabular}}
\end{table}
}

\subsubsection{Method Comparison}

We compare three methods with progressively richer information.

\noindent\textbf{Propagate All}: Predict YES for every pair. Due to the NL/PL boundary, we have no basis for distinguishing tainted from untainted returns. The only uninformed choices are to propagate all or propagate none—i.e., to treat every LLM return as either tainted or clean. This baseline adopts the conservative strategy of universal propagation. No LLM or taxonomy is required.

\noindent\textbf{LLM Code}:
Four models from independent families---GPT-5.5, Claude-Opus-4.8, DeepSeek-v4, and Qwen-3.7+~\cite{qwen2025}---predict from source code, prompt template, LLM output, and sink information, \emph{without} taxonomy labels. The prompt supplies five sections (source code, prompt template, LLM output, placeholder name/value, and sink location) and asks whether attacker-controlled placeholder content can propagate through the LLM response to the specified sink (full prompt in our artifacts repository). Using multiple families avoids circular evaluation.

\smallskip
\noindent\textbf{\toolname{}}:
The composability property of $R^\#$
(\S\ref{sec:transfer}) establishes that end-to-end
propagation from placeholder~$p$ to sink~$s$
decomposes into two independent factors:
$\mathrm{reach}(p)$ at the NL/PL boundary
(Def.~\ref{def:prop}) and code-level reachability
from the LLM output to the sink in the host program.
\toolname{} instantiates this decomposition.
Let $\mathcal{L}(p)$ denote the consensus taxonomy
labels assigned to placeholder~$p$, and let
$\mathcal{N}$ be the non-propagating set
(Def.~\ref{def:nonprop}). The taxonomy-induced
reachability summary $R^\#$
(Def.~\ref{def:transfer}) resolves the first factor:
when $\mathcal{L}(p) \subseteq \mathcal{N}$, the
placeholder is unreachable and no downstream analysis
is needed. For the remaining pairs,
$\mathrm{reach}(p)$ holds and the second factor is
resolved by a code-level analysis that receives only
source code and sink location, isolating code-level
reasoning from NL/PL-boundary semantics (details
in our artifacts repository):

\begin{equation}
\toolname{}(p, s) = \begin{cases}
  \textit{false}
    & \text{if } \mathcal{L}(p) \subseteq \mathcal{N}\;(R^\#\text{ unreachable}) \\
  \text{LLM}(p, s)
    & \text{otherwise}
\end{cases}
\label{eq:cv}
\end{equation}
Because the two factors are independent, the taxonomy
summary and code-level analysis are improved
or replaced separately. Our evaluation varies
code-level analysis across four model families to confirm this
independence (\S\ref{sec:robustness}). The reported
$F_1$ measures the composed predicate, not $R^\#$
alone.

\subsubsection{Comparison Results}

\autoref{tab:prediction} summarizes the results, ordered from baseline through single-signal methods to combined approaches. Propagate All achieves perfect recall but only 31.2\% precision, with 243 false positives making it unusable.
LLM Code reaches only 50.0--62.8\% $F_1$ across four model families: a single prompt must simultaneously infer what the LLM does with the placeholder \emph{and} trace code-level reachability, and does neither reliably.
\toolname{} addresses this by decomposing the problem: $R^\#$ resolves placeholder-level reachability via the taxonomy, letting the code-level analysis focus on a narrower and more tractable task.

{
\setlength{\abovecaptionskip}{3pt}
\begin{table}[t]
\caption{Propagation prediction on 353 pairs (110 YES / 243 NO) from 62 sink-containing files. LLM Code and \toolname{} are reported across model families for cross-family robustness; \toolname{} rows vary the code-level analysis model (all use 3-model consensus labels for the taxonomy filter).}
\label{tab:prediction}
\centering
\small
\begin{tabular}{@{}llrrr@{}}
\toprule
\textbf{Method} & \textbf{Model} & \textbf{Prec.} & \textbf{Rec.} & \textbf{$F_1$} \\
\midrule
Propagate All & --- & 31.2\% & 100\% & 47.5\% \\
\midrule
\multirow{4}{*}{LLM Code} & GPT-5.5 & 69.4\% & 39.1\% & 50.0\% \\
 & Opus-4.8 & 68.1\% & 44.5\% & 53.8\% \\
 & DeepSeek & 58.1\% & 68.2\% & 62.8\% \\
 & Qwen-3.7+ & 69.6\% & 53.9\% & 60.8\% \\
\midrule
\multirow{4}{*}{\textbf{\toolname{}}} & GPT-5.5 & 72.5\% & 93.6\% & \textbf{81.7\%} \\
 & Opus-4.8 & 72.8\% & 90.0\% & 80.5\% \\
 & DeepSeek & 69.7\% & 90.0\% & 78.6\% \\
 & Qwen-3.7+ & 73.5\% & 88.2\% & 80.2\% \\
\bottomrule
\end{tabular}
\end{table}
}

\subsubsection{Per-Label Security Analysis}

\autoref{tab:perlabel} reports the propagation rate for each label, i.e., the fraction of pairs with that label that are in the YES class. The discriminative power of the taxonomy comes primarily from the L0 (blocked) group.
When \emph{all} of a placeholder's consensus labels fall in $\mathcal{N}$, it rarely propagates to a sink.
Labels tied to executable output modality, specifically CLI Commands (87.5\%) and Code Snippet (85.2\%), show the highest propagation rates, consistent with the intuition that placeholders influencing code or command generation pose the greatest security risk. In practice, tool builders can assign elevated alert levels to these high-propagation labels, prioritizing review of code paths where LLM output is likely to reach executable sinks.

{
\setlength{\abovecaptionskip}{3pt}
\begin{table}[t]
\caption{Per-label propagation rates. Propagation rate = fraction of pairs with this label that are YES. Labels with $<$5 occurrences in the security dataset are omitted for stability.}
\label{tab:perlabel}
\centering
\setlength{\tabcolsep}{3pt}
\resizebox{0.48\textwidth}{!}{%
\begin{tabular}{@{}lrrr|lrrr@{}}
\toprule
\textbf{Label} & \textbf{Y} & \textbf{N} & \textbf{Rate} & \textbf{Label} & \textbf{Y} & \textbf{N} & \textbf{Rate} \\
\midrule
CLI Commands & 7 & 1 & 87.5\% & Fragment Copy & 40 & 109 & 26.8\% \\
Code Snippet & 52 & 9 & 85.2\% & Binary Decision & 4 & 16 & 20.0\% \\
Template Slotting & 43 & 57 & 43.0\% & Category Label & 4 & 22 & 15.4\% \\
JSON-Only Template & 23 & 31 & 42.6\% & Ignored & 4 & 26 & 13.3\% \\
Keyword Echo & 33 & 75 & 30.6\% & Paraphrase Rewrite & 6 & 48 & 11.1\% \\
Content Expansion & 55 & 144 & 27.6\% & Gen. Summarization & 0 & 11 & 0.0\% \\
\bottomrule
\end{tabular}}
\end{table}
}

\subsubsection{Error Analysis}
With the consensus labels, \toolname{} produces 17 false negatives and 17 false positives.
Most false negatives originate in the code-level analysis step: the taxonomy correctly identifies the placeholder as propagating, but the code-level analysis fails to trace indirect data flow through framework abstractions, callback chains, or multi-step pipelines.
The remainder are taxonomy labeling errors (e.g., a system prompt labeled \emph{Ignored} that nonetheless shapes generation style).
False positives arise when the code-level analysis predicts propagation but expert reviewers determined the sink is unreachable in practice.

\subsubsection{Robustness Analysis}
\label{sec:robustness}
To verify that the gains are not tied to the specific models used in labeling, we include another SOTA model, Qwen-3.7+, which was not involved in any
previous stage. Its $F_1$ of
80.2\% falls within the range of the other three
families (78.6--81.7\%), confirming that $R^\#$
generalizes across code-level analysis models.

\subsubsection{Cross-Language Validation on Real-World CVEs}

Because the taxonomy models LLM behavior rather than host-language syntax, it should generalize across programming languages. To validate this, we apply all three methods to real-world security vulnerabilities in OpenClaw, an open-source LLM agent framework written in TypeScript.
To construct the validation dataset, we systematically reviewed all closed prompt-injection vulnerabilities in the OpenClaw repository and attempted to reproduce each one on Claude Sonnet 4.5~\cite{claudesonnet45} and Gemini 2.5 Flash~\cite{gemini25flash}.
As the time of writing, 6 vulnerabilities remain reproducible, and the reminder have been mitigated by model-level safety training.
For each reproducible case, we extract both the vulnerable source code (before fix) and the patched source code (after fix) from the git history, yielding 12 (placeholder $\times$ sink) pairs: 6 YES (before-fix, vulnerable) and 6 NO (after-fix, fixed).
This dataset includes CVE-2026-27001 (CWD path injection, CVSS~8.6), CVE-2026-22175~\cite{cve202622175} (exec allowlist bypass, CVSS~7.1), and four additional vulnerabilities involving marker spoofing, Unicode bypass, and wrapper-fragment escape.

\begin{table}[t]
\setlength{\abovecaptionskip}{2pt}
\setlength{\belowcaptionskip}{-2pt}
\caption{Cross-language validation on 12 pairs from 6 real-world CVEs in OpenClaw. 6 YES (before-fix) / 6 NO (after-fix).}
\label{tab:openclaw}
\centering
\small
\begin{tabular}{@{}lrrrrr@{}}
\toprule
\textbf{Method} & \textbf{Prec.} & \textbf{Rec.} & \textbf{$F_1$} & \textbf{FP} & \textbf{FN} \\
\midrule
Propagate All & 50.0\% & 100\% & 66.7\% & 6 & 0 \\
LLM Code & 80.0\% & 66.7\% & 72.7\% & 1 & 2 \\
\textbf{\toolname{}} & \textbf{100\%} & \textbf{100\%} & \textbf{100\%} & \textbf{0} & \textbf{0} \\
\bottomrule
\end{tabular}
\end{table}

\autoref{tab:openclaw} reports the validation results. \toolname{} achieves perfect $F_1$ on this dataset, where the taxonomy labels precisely track the effect of each security fix. Consider CVE-2026-22175 (exec allowlist bypass): the attacker payload \texttt{busybox sh -c "cat /etc/shadow"} reaches the \texttt{spawn()} sink only if the model emits an \texttt{exec} tool call, a dependency invisible from the code. Before the fix, the model does so (label \emph{Content Expansion}, propagating); the patch causes it to refuse (label \emph{Policy Refusal}, non-propagating). Across all 6 cases, before-fix pairs are labeled \emph{Content Expansion} (the LLM uses injected content to generate actions), while after patching, 5 shift to \emph{Ignored} and 1 to \emph{Policy Refusal}, all non-propagating.
The fix fundamentally changes how the LLM processes attacker-controlled input: sanitization, marker isolation, or path validation causes the LLM to \emph{ignore} the injection, which are precisely captued by the non-propagating labels.
These results demonstrate that \toolname{} generalizes across languages (from Python to TypeScript) and vulnerability sources (from a curated dataset to real-world CVEs).

\begin{tcolorbox}[colback=gray!5,colframe=gray!50,boxrule=0.5pt,left=3pt,right=3pt,top=2pt,bottom=2pt]
\textbf{Answer to RQ2:}
\toolname{} reaches $F_1 = 81.7\%$, substantially outperforming both baselines and consistently improving performance across four model families (including a held-out model that never participated in labeling). On six real-world OpenClaw CVEs in TypeScript, it achieves a perfect $F_1$ score.

\end{tcolorbox}

\setlength{\abovecaptionskip}{3pt}
\begin{table}[t]
\caption{Backward slice reduction from taxonomy-informed barriers. Reduction = mean per-file ratio of lines removed.}
\label{tab:slicing}
\centering

\begin{tabular}{@{}lrr@{}}
\toprule
\textbf{Scope} & \textbf{Files} & \textbf{Mean Reduction} \\
\midrule
All files & 295 & 1.1\% \\
With barriers & 32 & 10.1\% \\
\textbf{With $>$0 cut} & \textbf{14} & \textbf{23.1\%} \\
\bottomrule
\end{tabular}
\end{table}

\subsection{RQ3: Backward Slicing with \toolname{}}
\label{sec:rq3}

Beyond the taint analysis for security, we evaluate the utility of \toolname{} for backward slicing.  

\subsubsection{Setup}
A backward slice from an LLM call's input arguments identifies all code lines (data and control dependencies) that influence the prompt sent to the LLM. Because the LLM is opaque, we cannot trace backward through it directly. Instead, the reachability summary $R^\#$ bridges this gap by identifying which input placeholders actually influence the output.
Traditional slicers treat the LLM call as opaque and include \emph{all} upstream dependencies indiscriminately, even those feeding placeholders that the LLM ignores.
The taxonomy provides the missing information: if a placeholder is labeled entirely with non-propagating labels ($\mathcal{L}(p) \subseteq \mathcal{N}$), its upstream code does not influence the LLM output and can be excluded from the slice.

We implement this using CodeQL~\cite{avgustinov2016ql} on 295 Python files, the subset of files that contain LLM callsites and for
which CodeQL successfully constructed an analysis.
CodeQL handles both data-flow dependencies (via taint tracking) and control-flow dependencies (via \texttt{getParentNode+()}, which captures enclosing \texttt{if}/\texttt{for}/\texttt{while} conditions).

For each file, we compute two slices from LLM call arguments:
\begin{itemize}[leftmargin=*]
\item \textbf{Full Slice}: standard backward slice including all placeholders' upstream chains (i.e., data flow via CodeQL taint tracking, and control flow via CodeQL's \texttt{getParentNode+()} for enclosing \texttt{if}/\texttt{for}/\texttt{while} conditions).
\item \textbf{\toolname{}-Informed Slice}: same query, but with CodeQL \texttt{isBarrier} predicates that block taint propagation through non-propagating placeholder variables. Barriers are per-file (variable, file) pairs: a variable is barriered in a given file only if it appears exclusively in non-propagating placeholders and never in propagating ones within that file.
\end{itemize}

From the 295 files, 32 contain at least one non-propagating placeholder variable, yielding 38 barrier variables total.

\subsubsection{Results}

\autoref{tab:slicing} reports the slice reduction. The global mean reduction (1.1\%) is modest because 89\% of files (263/295) have only propagating placeholders. The taxonomy correctly identifies nothing to cut.
Among the 32 files with at least one non-propagating variable, the mean rises to 10.1\%.
Focusing on the 14 files where the taxonomy actually removes lines, the \textbf{mean per-file reduction is 23.1\%}, with peaks reaching 65.4\% and 57.1\% shown in \autoref{tab:slicing_top}.

\subsubsection{Soundness Check}

To verify that removed lines do not sacrifice relevant dependencies, we check against the human-annotated ground truth from RQ1 (\S\ref{sec:rq1}).
Barriers are placed only on variables whose placeholders are exclusively labeled as non-propagating.
To confirm, we manually inspected all 70 cut lines across the 14 affected files and verified that none of them contribute content to the LLM output: every cut line belongs to the upstream chain of a placeholder that human annotators independently labeled as non-propagating (e.g. unused configuration variables).
The taxonomy-informed slice therefore achieves a 23\% reduction without sacrificing any ground-truth dependency.

{
\setlength{\abovecaptionskip}{3pt}
\setlength{\belowcaptionskip}{-5pt}
\begin{table}[t]
\caption{Top files by slice reduction. Full = full-slice lines; Prop = taxonomy-informed slice lines.}
\label{tab:slicing_top}
\centering
\small
\begin{tabular}{@{}lrrrr@{}}
\toprule
\textbf{File (abbreviated)} & \textbf{Full} & \textbf{Prop} & \textbf{Cut} & \textbf{Reduction} \\
\midrule
scratchTHOUGHTS/ripYT.py & 26 & 9 & 17 & 65.4\% \\
referral-augment & 7 & 3 & 4 & 57.1\% \\
langchain-tutorial/chat-memory & 23 & 11 & 12 & 52.2\% \\
codeGPT/analysis\_repo.py & 30 & 17 & 13 & 43.3\% \\
langroid/test\_llm.py & 13 & 9 & 4 & 30.8\% \\
instructor/simple\_prediction & 7 & 5 & 2 & 28.6\% \\
\bottomrule
\end{tabular}
\end{table}
}

\autoref{tab:slicing_top} details the top cases. In these files, non-propagating placeholders (system prompt constants, configuration variables, and agent state) have non-trivial upstream definition chains involving imports, loop state, and file I/O that the traditional slicer unnecessarily includes.

\subsubsection{Why Reduction Is Bounded}

Three structural factors limit the achievable reduction:
(1)~Non-propagating and propagating placeholder variables frequently \emph{merge} at the prompt construction site (e.g., \texttt{prompt = f"\{SYSTEM\} \{user\_query\}"}), so the merged line and shared downstream code cannot be cut.
(2)~Many non-propagating variables have only 1--2 upstream lines (a constant definition or import), limiting the per-variable savings.
(3)~In most LLM-integrated code, the majority of placeholders carry user content and are propagating. Files dominated by system prompts and configuration variables are the minority.

Despite these limits, the taxonomy provides slicing tools with LLM-aware semantic information that was previously unavailable.
No prior backward slicer can distinguish a placeholder the LLM uses from one it ignores.

\begin{tcolorbox}[colback=gray!5,colframe=gray!50,boxrule=0.5pt,left=3pt,right=3pt,top=2pt,bottom=2pt]
\textbf{Answer to RQ3:}
\toolname{} also helps improve backward slicing. For non-propagating placeholders that carry upstream code, \toolname{} reduces the backward slice by an average of 23\% without eliminating any true dependencies.
\end{tcolorbox}

\section{Discussion}
\label{sec:discussion}

\subsubsection*{Information contribution and error attribution}
Before \toolname{}, no analysis could determine whether the content of a given placeholder actually manifests in the LLM output. Therefore, the downstream analyses faced the same binary choice: either assume all placeholders propagate, or assume none do.
\toolname{} resolves this by providing per-placeholder reachability information ($R^\#$) that was previously unavailable. Equipped with this information, downstream analyses can focus their budget on the placeholders that actually matter, which is why \toolname{} achieves $F_1 = 81.7\%$ while Propagate All and LLM Code plateau at 47.5\% and 62.8\% respectively.
The remaining errors do not stem from incorrect reachability taxonomy. In the vast majority of cases, \toolname{} correctly identifies the placeholders whose content propagates to the LLM output. Instead, the residual loss in precision arises from downstream code-level reasoning, which is orthogonal to \toolname{}'s contribution and can be improved through independent advances in program analysis.

\subsubsection*{Limitations}
The primary limitation of \toolname{} is that $R^\#$ provides soundness only \emph{relative to the labeling} $\mathcal{L}$ (\S\ref{sec:transfer}). If all labels assigned to a genuinely reachable placeholder fall inside $\mathcal{N}$, the summary under-approximates. A classical soundness proof would require a formal abstraction of $M$ (e.g., via Galois connections~\cite{cousot1977abstract}), but this is precluded by the LLM's opacity (\textbf{C1}). Instead, we bound the failure rate empirically: cross-family consensus labeling achieves $\kappa = 0.77$ (\S\ref{sec:rq1}), Good--Turing estimation places the probability of encountering an unlabeled behavior at 0.09\%, and human--model agreement reaches $\kappa = 0.72$. These constitute the strongest soundness guarantee available without white-box access to $M$.
Additionally, our current evaluation applies $R^\#$ to individual LLM callsites. Programs with chained LLM calls (where the output of one call feeds into the prompt of another) can be handled by applying the same per-callsite labeling procedure at each boundary in sequence, but we leave this compositional extension to future engineering work.

\subsubsection*{Threats to Validity}
\label{sec:threats}

Our dataset consists of Python applications from open-source GitHub repositories that use LLM calls (including OpenAI and LangChain). Consequently, the results may not fully generalize to other programming languages, closed-source applications, or alternative LLM providers. We mitigate these threats in several ways. First, we evaluate \toolname{} on the OpenClaw benchmark (\S\ref{sec:rq2}), providing initial cross-language evidence on TypeScript. Second, our approach treats the LLM as a black box and depends only on prompt-response behavior rather than provider-specific internals, making it applicable in principle to other LLM providers. Nevertheless, our evaluation does not include proprietary codebases, which we remain for future work.

\section{Related Work}
\label{sec:related}

\subsubsection*{Dataflow analysis across opaque boundaries}
Prior work has constructed dataflow summaries for several opaque boundaries (\S\ref{sec:reachability}), including HTTP~\cite{tripp2009taj}, Android lifecycle~\cite{arzt2014flowdroid,li2015iccta}, JNI~\cite{enck2010taintdroid}, cross-language calls~\cite{li2022polycruise}, and distributed systems~\cite{fu2021flowdist}.
On the slicing side, modern tools such as CodeQL~\cite{avgustinov2016ql} use barrier predicates to prune irrelevant flows, and recent advances apply graph simplification to accelerate path-sensitive analysis~\cite{cheng2024fgs}.
All of these techniques assume a deterministic, code-level input-output mapping at the boundary. None provides a dataflow summary for the NL/PL boundary, where the transformation is mediated by natural language and governed by an opaque model.

\subsubsection*{Security analysis of LLM-integrated systems}
Prompt injection~\cite{greshake2023not,schulhoff2023hackaprompt,liu2023prompt} and jailbreaking~\cite{wei2023jailbroken} characterize attack vectors against LLMs. B{\"o}hme et al.~\cite{bohme2024security} identify opaque model boundaries as a key open challenge for software security.
On the analysis side, Fides~\cite{costa2025securing} tracks confidentiality labels at runtime in LLM agent planners, AgentFuzz~\cite{agentfuzz2025} fuzzes LLM agents to discover taint-style vulnerabilities, IRIS~\cite{li2025iris} uses LLMs to infer taint specifications for traditional code, and LATTE~\cite{latte2025} automates binary taint analysis with LLM-identified sources and sinks.
Concurrently, TaintP2X~\cite{he2026taintp2x} addresses the output-to-sink leg: it models LLM-generated outputs as taint sources and tracks their propagation to sensitive sinks via static analysis with LLM-assisted false-positive pruning.
\toolname{} addresses the complementary input-to-output leg: which placeholders actually manifest in the LLM output. The two operate at different boundaries.

\section{Conclusion}
\label{sec:conclusion}
LLM API calls create the NL/PL boundary, which breaks the program analysis relying on dataflow summaries. We address this gap with \toolname{}, the first reachability model for the NL/PL boundary. Built from 8,119 real-world pairs, its 25-label taxonomy captures whether the LLM uses a given input. It does so by observing external behavior alone, discretizing the diverse transformations into a finite label space, and labeling per callsite. In our evaluation, \toolname{} significantly improves two downstream analyses. In taint analysis, \toolname{} achieves $F_1 = 81.7\%$, nearly double the conservative baseline. It also flags all vulnerable flows on six real-world OpenClaw CVEs, even in TypeScript. In backward slicing, the same labels prune roughly a quarter of the slice without losing a real dependency. Because both analyses build on the same primitive, \toolname{} has the potential to benefit a wide range of program analyses at the NL/PL boundary.

\section*{Data Availability and Acknowledgments}
Our dataset is available at \url{https://anonymous.4open.science/r/nlpl-boundary-artifacts-6DBB/README.md}.
This work uses generative AI tools (GPT-5.2/5.5, Claude-Opus-4.8, DeepSeek-v4, Qwen-3.7+) for callsite reconstruction, output generation, taxonomy labeling, and propagation verification as described in the paper, and polishes the wording.

\newpage

\bibliographystyle{IEEEtran}
\bibliography{references}

\end{document}